\documentclass[doublecol]{epl2}
\usepackage{float}
\title{The Earth's Frame Dragging via Laser Ranged Satellites: a Response to ``Some considerations
 on the present-day results for the detection of frame-dragging after the final outcome of
 GP-B''
by L. Iorio.}
\shorttitle{} 

\author{J. C. Ries\inst{1} \and I. Ciufolini\inst{2,3} \and E. C. Pavlis\inst{4} \and A. Paolozzi\inst{5}  \and  R. Koenig\inst{6} \and R. A. Matzner\inst{7} \and G. Sindoni\inst{5} \and H. Neumayer\inst{6}}
\shortauthor{J. C. Ries \etal}

\institute{
 \inst{1} Center for Space Research, University of Texas at Austin, USA \\
  \inst{2} Dip. Ingegneria dell'Innovazione, Universit\`a del Salento, Italy \\
  \inst{3}  INFN Sezione di Lecce, Italy \\
  \inst{4} Joint Center for Earth Systems Technology, University of Maryland, Baltimore County, USA \\
  \inst{5} Scuola di Ingegneria  Aerospaziale, Sapienza Universit\`a di Roma Italy \\
  \inst{6} GFZ German Research Centre for Geosciences, Potsdam, Germany \\
  \inst{7} Center for Relativity, University of Texas at Austin, USA \\
   }
   \pacs{0.4}{First pacs description}

\abstract{ In this note, we reply to the preceding paper by Iorio
\cite{bib9}, hereafter referred to as I2011, where we address
criticisms regarding the Lense-Thirring
frame-dragging experiment results obtained from the laser ranging to
the two LAGEOS satellites.}

\begin{document}

\maketitle

\section{Introduction}

To put this discussion into context, a short summary of the
developments to this point is needed. In the late 1980s, it was
proposed to launch a satellite (LAGEOS-3) into a supplementary orbit
to the existing LAGEOS satellite (LAGEOS-1) in order to separate the
secular orbit node precession caused by frame-dragging (the
Lense-Thirring effect predicted from General Relativity) from the
much larger but similar effect due to the even zonal geopotential
harmonics ($J_{n}$, where n is even) \cite{bib_ciuf}. This orbit
configuration would cancel out the influence of the errors in the
even zonals, while the frame-dragging effect would be the same for
both and thus still remain. It was understood that the orbit
injection would only need to be accurate enough to cancel the
uncertainty in the even zonal effects \cite{bib1}, not the total
effect of the even zonals on the orbit. We also emphasize that even
at that time, the issue of whether laser ranging could measure the
frame-dragging effect was not an issue. By then, the models and data
were already more than adequate to observe this signal. The problem
was simply separating that signal from a similar, but significantly
larger, signal caused by errors in the even zonals. This experiment
was considered important enough that it warranted a joint NASA/ASI
study to evaluate the feasibility of the result \cite{tap89}. The
primary implementation issue was related to the cost of the launch
vehicle to place the satellite in the required
orbit.\\
\\
 Although LAGEOS-3 was not selected as a NASA mission, the
determination of the Earth's gravity field continued to improve, and
a second LAGEOS satellite (LAGEOS-2) was launched in late 1992,
though not into the inclination that would have been best for the
frame dragging experiment. In 1998, a test of frame-dragging with an
accuracy on the order of 20\% was announced \cite{bib2}. In order to
remove the effect of the two largest even zonal errors (due to
$J_{2}$ and $J_{4}$), this result relied on a linear combination of
the residual node rates of LAGEOS-1 and LAGEOS-2 as well as the
perigee of LAGEOS-2. Frame-dragging also affects the orbit perigee
in a secular way, as do the even zonals, and a particular linear
combination was used to cancel the two largest errors. However,
there were significant concerns with this experiment, most notably
due to the use of the uncertain perigee signal and the true accuracy
of the gravity field available at the time  \cite{bib14}.\\
\\
Around this time, NASA was preparing to launch a joint gravity
mission with the German space agency. Based on preliminary results
of the improved gravity fields from GRACE \cite{bib17}, an accurate
test of frame-dragging would be possible using only the node signals
from the two LAGEOS satellites \cite{bib_pav, bib13}. Soon after,
such a result was announced \cite{bib3} and confirmed independently
a few years later \cite{bib15}. A complete description of the
combined analyses is available in \cite{bib5,bib4,bib_ciuf11}. Our
joint conclusion is that the GRACE-based gravity models are indeed
accurate enough that the General Relativity prediction of
frame-dragging is confirmed with an error of approximately 10\%.

\section{Soundness of the Analysis Method}
We address the critique in I2011 of the analysis method first. The
author indicates that the analysis of geodetic signals in time
series of orbit element residuals is inadequate, and states that the
frame dragging effect must be ``explicitly modeled and solved-for in
the data reduction process.'' This ignores the value of decades of
analyses that have successfully used orbit element residuals to
determine geodetic parameters, dating back to the Vanguard satellite
whose orbit eccentricity variations provided the first estimate of
the Earth's `pear shape' \cite{bib12}. Today, this technique has
been developed into the more sophisticated orbit element excitation
approach \cite{bib6, bib10}. Explicit parameter estimation is of
course widely used, particularly when large numbers of parameters
have to be estimated or an error covariance matrix is essential, but
analysis of orbit element residuals has significant advantages as
well \cite{bib6}. The ability to plot the time series and see how a
predicted mechanism fits provides invaluable insight into the
problem that parameter estimation cannot provide. Our approach
provides a valuable visual confirmation, illustrated so well in
Figure 1, that the unmodeled variations in J$_{2}$ are indeed
canceled in the J$_{2}$-error-free combination time series. The time
series can be band-limited filtered and outliers can be detected and
edited, and all of this can be accomplished with little
computational expense. Finally, simply estimating a parameter and
looking at the covariance does not account for the effect of the
systematic errors that typically dominate the error. As in the case
for the original LAGEOS-3 analysis \cite{tap89, asi}, where the
frame-dragging parameter was explicitly estimated, a post-fit error
analysis was still required to obtain a realistic uncertainty
assessment. The formal errors are often so overly optimistic that
they provide no useful uncertainty information, particularly when an
experiment is dominated by a few systematic errors, as is the case
here.\\
\\
 In the case of the LAGEOS frame-dragging experiment, a
particularly simple but elegant approach is used. The basic time
series  used for our analysis is the sequence of the orbit end-point
overlaps, which are the mismatch between the end of one arc and the
beginning of the next arc. These are converted into orbit node
residuals that are then accumulated over time to generate the
long-term node drift signal for each satellite, as illustrated in
Figure 1. These two node signals are combined by multiplying the
LAGEOS-2 residual node drift by a constant c and adding it to the
node signal from LAGEOS-1. By choosing a specific value for c, based
on each satellite's orbit inclination and altitude, the largest
error due to $J_{2}$ can be eliminated. This $J_{2}$-error-free
combination, also shown in Figure 1, is compared to the same linear
combination of the node precessions predicted by General Relativity.
In addition to our own internal checks (described later), the use of
orbit overlaps to determine a residual node time series was
determined through independent analysis  to work correctly for
secular signals \cite{bib11}. So we find it curious that the author,
while criticizing our use of this method, has applied the same
method for his own frame-dragging test using the Mars Global
Surveyor (MGS) orbit overlaps \cite{bib7,bib8}. In \cite{bib8}, the
author notes that ``effects like the Lense-Thirring one,
accumulating in time, are, instead, singled out'' and refers to the
same \cite{bib11} reference; these MGS results of the author
(implying a 6\% confirmation of General Relativity) are particularly
surprising in light of there being no dedicated gravity mission at
Mars to provide the same kind of high-accuracy gravity field model
that makes the Earth-based experiment possible at the 10\% level. In
any case, we conclude that the author's concern over the use of
orbit element overlaps to observe the drift in the node due to the
frame-dragging effect is unfounded.

\section{Residual Analysis}
Returning to the analysis method, the best available modeling is
used in a least-squares fit to the laser ranging data for a
continuous dynamical orbit based on the numerical integration of a
complex dynamical model starting with a given set of initial
conditions. There will always be some modeling error that remains.
In our case, this is dominated by the even zonal errors and not
modeling (by choice) the frame dragging effect. The least-squares
orbit fit minimizes the variance of the laser range residuals. As
long as no empirical parameters are estimated that can accommodate
the errors in the out-of-plane orbit components (node and
inclination), these zonal errors and unmodeled  frame dragging
effect must be reflected in the residuals. The author of I2011
affirms this in \cite{bib7} by noting that the frame-dragging
effect, by not being modeled in the MGS orbit fits, ``is not
included, so that it is fully accounted for by the showed
residuals.'' In the least-squares fit to the range data, only the
mean orbit error can be accommodated (via the adjustment of the
orbit initial conditions). This adjustment of the mean orbit
minimizes the overall variance, but results in approximately equal
residuals near the ends of the arc and smaller residuals near the
arc mid-point, as this point is where the mean (but slightly
incorrect) orbit tends to be closest to the true orbit. The result
is the well-known `bow-tie' in the orbit error \cite{bib16}, which
can be clearly seen in Figure 2, where we show the actual residuals
for a typical 14-day fit. We note that in I2011, the author
complains that no such residual plots were ever shown, but a plot of
tens of thousands of residuals would have provided no useful
information. Of course we gladly provide the residuals for every
laser range observation if asked. Also we note that the residuals in
Figure 2 are not inconsistent with Fig. 5 in I2011, after taking
into account the effect of the initial conditions adjustment and
that the residuals must also include the effect of the even zonal
errors. The author's concern about the LAGEOS residuals not properly
reflecting the modeling errors (including not modeling the
frame-dragging effect) is
unfounded.\\
\\
 For the best orbit fits, special empirical acceleration parameters
are typically estimated which are highly effective in removing what
is essentially a slow drift in the orbit elements (both in-plane and
out-of-plane) over the fit interval. It is these parameters that
allow the determination of sub-cm-level orbits for the LAGEOS
satellites in spite of imperfect force models. However, for this
experiment, the cross-track (out-of-plane) accelerations are not
estimated to avoid absorbing the unmodeled frame-dragging effect on
the node. This also leaves in the residuals the effect due to the
errors in the even zonals, with the result shown in Figure 2. These
fits are not sub-cm, but rather closer to 3-4 cm (for data that is
precise at the few-mm level for the best stations). The in-plane
accelerations are estimated to accommodate the significant
along-track non-gravitational modeling errors, but this has no
effect on the cross-track (out-of-plane) component. This is derived
from the well-known Lagrange Planetary Equations (in Gauss's form),
which clearly indicate that the forcing in the along-track direction
has no direct effect on the orbit node. This is, in fact, recognized
 by the author himself, where in \cite{bib7}, he states
that because only along-track (in-plane) empirical accelerations
were estimated for MGS, ``it is unlikely that the out-of-plane
Lense-Thirring signal was removed from the data in the least-squares
procedure.'' Furthermore, in the independent tests performed by Ries
\cite{bib15}, certain internal checks were conducted. In one case,
the frame-dragging effect was modeled, and the change in the
residual time series matched exactly the relativity prediction. This
is a very strong internal verification; if the use of empirical
in-plane parameters was in some way absorbing or creating the
frame-dragging signal, a perfect shift in the resulting time series
is not possible. A second test where the geodesic precession was
also not modeled resulted in precisely a 57\% change in the
frame-dragging signal, which is exactly what would be expected since
the component of the geodetic precession in the equatorial plane is
57\% of the frame-dragging effect. Consequently, the author's
concern that the frame-dragging effect (on the node) has been
accommodated or distorted in some way due to the estimation of
in-plane empirical parameters is unfounded.

\section{The J$_{2}$-Free Linear Combination}
The author of I2011 devotes considerable discussion to the need to
compute the combination coefficient c with exquisite precision, but
the argument presented is incorrect. Recall that the two node
signals are combined by multiplying the LAGEOS-2 residual node drift
by a constant c and adding it to the residual node drift from
LAGEOS-1, where c=0.545 was used by Ciufolini et al.
\cite{bib5,bib4, bib_ciuf11}. This produces a J$_{2}$-error-free
node drift signal that is then compared to the same linear
combination of the frame-dragging signal predicted by General
Relativity. This is the critical point. The two linear combinations
are compared to each other, and the level of agreement reflects the
closeness in agreement to General Relativity. Referring back to the LAGEOS-3
experiment, the orbit inclination only had to be precise enough to
cancel the \underline{errors} in the gravity model, \underline{not
the full signal}. The same is true here, as can be seen by
considering the following. If we suppose that the errors in the
gravity field are so small that they are negligible, then the two
residual node time series will contain only the frame-dragging
signal. In that case, any linear combination of the two nodes will
match the same linear combination of the General Relativity predictions for the two
satellites. The value of c becomes arbitrary. It should be clear that
the combination constant c only needs to be known well enough
compared to the residual gravity modeling errors. In the case of our
~10\% experiment, c needs to be known with an accuracy of only a few
percent. Even considering the range between c=0.545 used in
Ciufolini et al. \cite{bib5,bib4, bib_ciuf11} and the value of
0.54097 proposed by I2011, the impact on the result is only about
2\%. Recomputing c on an arc-by-arc basis to more closely reflect
the evolving orbit parameters affects the result by only 1\%. This
also renders irrelevant the concern about the knowledge of the
orientation of the Earth's spin axis; it is known with an accuracy
that exceeds the requirement for calculating c to a few digits. We
conclude that the author's concern about requiring ultra-precise
knowledge of the combination factor c is unfounded.
\section{Reference Frame Effects}
The author of I2011 raises a vague concern that the geocentric
reference frame used for the satellite data analysis is somehow
determined in a favorable way for this experiment, since the LAGEOS
data are used as part of the reference frame definition. However,
what is primarily relevant for the laser ranging experiment is the
rotation of the node with respect to inertial space, since this
represents the absolute relativistic signal of interest. This is why
it is necessary to model the geodesic precession since this is a
similarly appearing component in the node precession, and we take as
given that geodesic precession has been validated by independent
tests. While it is difficult to imagine how the use of laser ranging
data for the reference frame could favorably `imprint' the
frame-dragging signal in these results, it is essentially impossible
in any case. Earth rotation (often called UT1), over long time
scales, is entirely determined by Very Long Baseline Interferometry
to distant quasars. Polar motion is currently determined almost
entirely by VLBI and Global Positioning System data. Furthermore,
polar motion is the crust-fixed motion of the Earth's rotation axis.
As a consequence, it is modulated by the Earth's rotation and cannot
look like a drift with respect to inertial space, which is the only
signal that would affect the LAGEOS experiment. It is noted that the
independent analysis by Ries et al. \cite{bib15} did not estimate
polar motion corrections, but relied entirely on Earth orientation
from other techniques and obtained a similar experimental result.
The concerns in I2011 about the impact of laser ranging on the
reference frame, resulting somehow in a favorable experiment
outcome, are thus unfounded.

\section{Closing Remarks}
Finally, we address some of the material in the introduction of
I2011. The author notes the cost of the GP-B mission, that it lasted
for 52 years (data was actually collected for ~12 months only), that
the mission was designed especially for a test of General
Relativity, and that there are some comments in the literature
questioning the investment in that mission. We ask how any of this
is relevant to the reliability of the results from laser ranging,
and we are a bit surprised about these clearly prejudicial remarks.
Further, the remark by the author that
the LAGEOS satellites ``were originally launched for different
purposes'' seems to be a way to imply, without actually saying it,
that the satellites are not well suited for this experiment. On the
contrary, the two satellites are nearly ideal test objects. The
LAGEOS satellites, being solid metal, have very low area-to-mass
ratios, which strongly attenuates the surface force effects on the
orbits. The satellites are high enough to attenuate the effect of
the errors in all but the lowest degree zonal harmonics, but not so
high that the frame-dragging signal is not clearly observable. The
only way to make the experiment better would be to have launched
LAGEOS-2 into the original LAGEOS-3 supplementary orbit.


\begin{figure}[h]
\onefigure{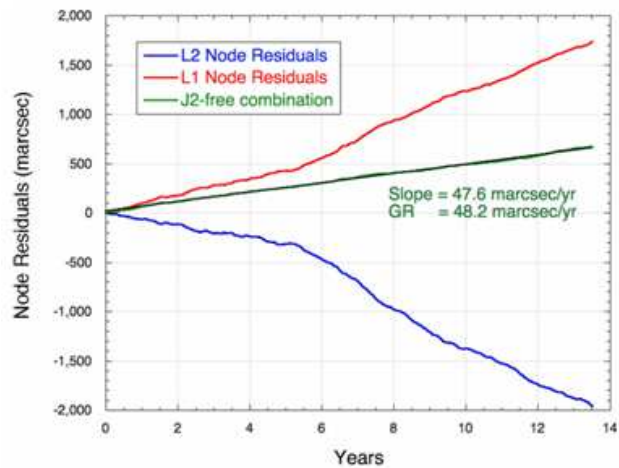} \caption{Residuals of the node of LAGEOS-1
(in red), the node of LAGEOS-2 (in blue), and the J2-free linear
combination of the two nodes (in green). The effect of the J2 error
(which dominates the two node residual trends) is larger for
LAGEOS-2 because of its lower inclination. The linear combination,
based on each satellite's altitude and inclination, removes the
specific effect of all errors in J2, including its unmodeled decadal
variations. Figure from Ries et al. \cite{bib15}.}
 \label{fig.1}
\end{figure}

\begin{figure}[H]
\onefigure{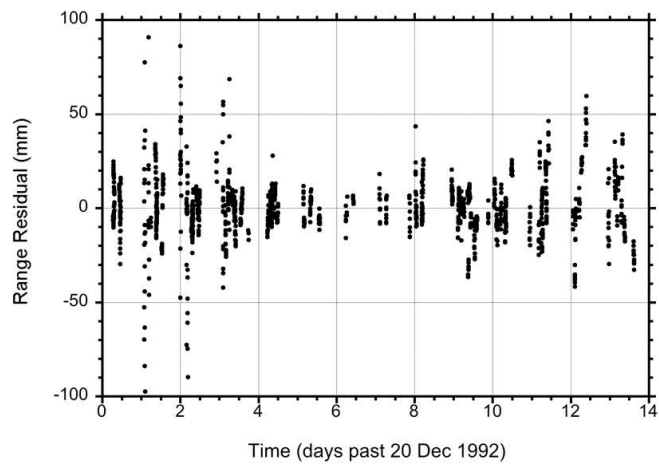} \caption{Laser range residuals for a
typical 14-day fit for LAGEOS-1. Because no cross-track empirical
accelerations are estimated, the drift in the node (caused by errors
in the even zonals and not modeling frame-dragging) over the fit
interval cannot be accommodated. The fit orbit differs from the true
orbit by the node drift, and this mismatch is reflected in the
residuals as a bow-tie shaped modulation of the cross-track orbit
errors. It is also reflected in the orbit end-point overlaps, which
are used to construct the residual node time series. }
 \label{fig.2}
\end{figure}


\acknowledgments


\begin{thebibliography}{0}

\bibitem{bib9}  
  \Name{Iorio L.}
  \REVIEW{Some considerations on the present-day results for the detection of frame-dragging after the final outcome of GP-B,  Europhysics Letters}{}{2011}{}.


\bibitem{bib_ciuf}
 \Name{Ciufolini I.}
 \REVIEW{Measurement of the Lense-Thirring drag on
high-altitude laser-ranged artificial satellites. Phys. Rev.
Lett.}{56} {1986}{278-281}.

\bibitem{bib1}
  \Name{Casotto S.}
  \REVIEW{Orbit injection error analysis for the proposed LAGEOS-3 mission, Celestial Mechanics and Dynamical Astronomy}{56}{1993}{397-408}.


    \bibitem{tap89}
\Name{Tapley B., Ries J.C., Eanes R.J. and Watkins M.M.}
\REVIEW{NASA-ASI Study on LAGEOS III, CSR-UT publication. Austin,
Texas CSR-89-3}{}{1989}{}






  \bibitem{bib2}   
  \Name{Ciufolini I., Pavlis E. C.,  Chieppa F.,  Fernandes-Vieira E. \and Perez- Mercader J.}
  \REVIEW{Test of General Relativity and Measurement of the Lense-Thirring Effect with Two Earth Satellites, Science}{279}{1998}{2100-2103}.

%

\bibitem{bib14}   
  \Name{Ries J.,Eanes  R.  \and  Tapley B.}
\Book{Nonlinear Gravitodynamics. The Lense-Thirring Effect}
 \Editor{R. Ruffini and C. Sigismondi }
  \Publ{World Scientific, Singapore}
  \Year{2003}
  \Page{201-211}.

\bibitem{bib17}
 \Name{ Tapley B., Bettadpur S., Watkins M. \and  Reigber Ch.}
  \REVIEW{The GRACE Gravity Mission: Status and Early Results, Geophys. Res. Lett.} {31}{2004}{L09607}.


\bibitem{bib_pav}
\Name{Pavlis E. C.} \Book{ Recent Developments in General
Relativity}
 \Editor{R. Cianci, R. Collina, M. Francaviglia, P. Fr\'{e} }
 \Year{2000}
  \Publ{Springer-Verlag, Milan}
  \Page{217-233}

\bibitem{bib13}   
  \Name{Ries J., Eanes R., Tapley  B. \and Peterson G. E.}
  \REVIEW{Prospects for an Improved Lense-Thirring Test with SLR and the GRACE Gravity Mission, proc. 13th International Laser Ranging Workshop, Washington, D.C., October 7-11, 2002 NASA/CP-2003-212248}{}{2003}{67-73}.

\bibitem{bib3}  
  \Name{Ciufolini I. \and Pavlis E.}
  \REVIEW{A confirmation of the general relativistic prediction of the Lense-Thirring effect, Nature}{431}{2004}{958-960}.


\bibitem{bib15}  
  \Name{Ries J., Eanes R. \and  Watkins M.}
  \REVIEW{Confirming the Frame-Dragging Effect with Satellite Laser Ranging,16th International Laser Ranging Workshop  October 12–17 2008, Poznan, Poland}{}{2008} {}


\bibitem{bib5}   
  \Name{Ciufolini I., Paolozzi A.,  Pavlis E.,  Ries J.,  Koenig R.,  Matzner R., Sindoni G. \and Neumayer H.}
  \REVIEW{Towards a One Percent Measurement of Frame-dragging by Spin with Satellite Laser Ranging to LAGEOS, LAGEOS 2 and LARES and GRACE Gravity Models, Space Science Reviews}{148}{2009}{71-104}.


\bibitem{bib4}   
  \Name{Ciufolini I., Paolozzi A., Pavlis E., Ries J.,  Koenig R., Matzner  R. \and  Sindoni G.}
  \Book{General Relativity and John Archibald Wheeler}
  \Editor{I. Ciufolini and R. A. Matzner }
  \Publ{Springer Verlag, Dordrecht}
  \Year{2010}
  \Page{371-434}.

\bibitem{bib_ciuf11}
\Name{Ciufolini I.,  Paolozzi A.,  Pavlis E., Ries  J., Koenig R.,
 Matzner R.,  Sindoni G. \and Neumayer H. }
 \REVIEW{Testing gravitational
physics with satellite laser ranging, The European Physical Journal
Plus} {126} {2011}{72}

\bibitem{bib12}  
  \Name{O'Keefe J.}
  \REVIEW{North-South asymmetry of the Earth's figure, Science}{130}{1959}{978-979}.

 \bibitem{bib6}  
  \Name{Eanes R.}
  \REVIEW{A study of temporal variations in Earth's gravitational
field using LAGEOS-1 laser range observations, CSR-UT publication,
Austin, Texas CSR-95-7}{}{1997}{}

\bibitem{bib10}  
  \Name{Lucchesi D.}
  \REVIEW{Reassessment of the error modeling of non-gravitational perturbations on LAGEOS II and there impact on the Lense-Thirring determination. Part 1, Planet. Space Sci.}{49}{2001}{447-463}.

\bibitem{asi}
\Name{Ciufolini I. et al.} \REVIEW{ASI-NASA Study on LAGEOS III,
CNR, Rome, Italy}{} {1989}

\bibitem{bib11}
  \Name{Lucchesi D. \and Balmino G.}
  \REVIEW{Planet. Space Sci}{54}{2006}{581-593}.



\bibitem{bib7}  
  \Name{Iorio L.}
  \REVIEW{Evidence of the gravitomagnetic field of Mars, Class. Quantum Grav.  }{23}{2006}{5451-5454}.

\bibitem{bib8}  
  \Name{Iorio L.}
  \REVIEW{On the Lense-Thirring test with the Mars Global Surveyor in the gravitational field of Mars , Central European Journal of Physics}{8}{2010}{509-513}.


\bibitem{bib16}   
    \Name{Tapley B. \and  Ries J.}
  \Book{Encyclopedia of Space Science and Technology}

   \Publ{Wylie and Sons, Hoboken, New Jersey}
  \Year{2003}
  \Page{341-355}.













\end{thebibliography}
\end{document}